# Phase Field Modeling of Galvanic Corrosion

Anahita Imanian and Mehdi Amiri[1]

Technical Data Analysis Inc.,

3910 Fairview Park Drive, Falls Church, VA, 22042, USA

**Abstract:**

A phase field (PF) based electrochemical model is presented for simulation of galvanic corrosion. Distributions of electrolyte potential and current density on anode and cathode surfaces are obtained by coupling the phase field variable with electrochemistry. Evolution of surface recession is naturally obtained by solving the phase field equations without tracking the evolving boundary. Numerical implementation involves solving the governing equations on a fixed mesh. The sharp interface as the limit of the PF model is shown by an asymptotic analysis. Two benchmark problems are discussed: a magnesium alloy-mild steel couple exposed to 5% NaCl solution and crevice corrosion for nickel in 1N sulfuric acid. A comparison is made considering available experimental data as well as other simulation data by an Arbitrary Lagrangian-Eulerian (ALE) method. A good agreement is obtained.

1. **Introduction**

Galvanic corrosion, resulting from galvanic coupling of two dissimilar metals in a corrosive medium, is one of the most common types of corrosion. In practical applications, the galvanic corrosion may be realized at macro and micro scales depending on the scale of analysis. For example, for automotive applications, macro-galvanic corrosion of Magnesium (Mg) becomes crucial when coupled with either Al alloys or steels. At micro scale, galvanic corrosion of Mg may occur due to galvanic activity between its primary constituents namely, primary $α$, eutectic $α$ and $β$ phases [1]. Consequently, damage morphology may appear different at micro and macro scales. The macro-galvanic corrosion may manifest itself as surface recession with high severity near the junction while the micro-galvanic corrosion is highly localized in the region of $β$ phase and the surrounding Al-rich-$\alpha$ area [1]. Predictive models and computational tools help engineers and designers make good design and address repair issues by predicting and quantifying galvanic corrosion at material interfaces. For simplicity of computational analysis, it is commonly assumed that only anodic sites corrode. Numerical simulation of galvanic corrosion, at both micro and macro scales, involves tracking the anode-electrolyte interface in the computational domain. Evolution of corroding surface makes the galvanic corrosion a moving boundary problem with appropriate boundary condition at the moving interface [2,3]. This is reminiscent of the Stefan problem [4], although the boundary conditions might be different. The interface between electrolyte and solid surface must be tracked at each computational time step so the boundary conditions can be imposed. Such interfaces impose discontinuity in the computational domain. For example, in corrosion, interface between electrolyte and solid material is referred to as discontinuity.

---

[1] Corresponding author: Email: mamiri@tda-i.com (M. Amiri), Tel: +1 703-237-1300.

Traditional approaches for solving such problems use moving mesh or re-meshing techniques [5–7]. In the moving mesh technique, meshes should be aligned with the moving interface in every time-step which increases computational cost particularly for complex geometries. In the re-meshing technique, on the other hand, new meshes are generated from mapping the old mesh variables along the moving interface at each time-step which may compromise the accuracy. Efficient alternatives are nodal enrichment strategies such as extended finite element methods (X-FEM) which improve the finite element shape functions with discontinuous fields and can be used effectively for modeling complex geometries [8]. Level set method [9,10] is another method for solving problems with moving interface in that it is independent of the interface geometry. In the level set method the interface is represented by a signed distance function. Value of the distance function at any point indicates the signed distance from that point to the moving interface. For example, the positive value of this function at a point implies that the point locates in direction of interface motion. The level set method is particularly useful for identifying enriched nodes and computing enriched matrices in the X-FEM [11]. Combining numerical advantages that level set and X-FEM methods provide, some moving interface problems such as pit growth, phase change and crack growth can be tackled [12,13]. Although these numerical techniques have advantages over the standard finite element method for modeling arbitrary discontinuities, there are challenges in using them. For example, numerical convergence of these methods strongly depends on the mesh size when it tends to zero. There is also a lack of rigorous theory to predict the propagation direction of a discontinuity such as a new crack segment [14]. Furthermore, tracking of the complex interface patterns like crack branching and 3D cracks becomes hard to model [15].

In order to circumvent the difficulty of tracking the moving interfaces, mesh free methods such as Peridynamic have been used to predict corrosion [16]. Chen and Bobaru [16] developed a Perdynamic model for pitting corrosion in which the evolution of concentration of metal ions is formulated in terms of integral equations rather than partial differential equations. This way, the evolution of moving interface is obtained as part of solution, thus eliminating difficulties that are inherent in traditional moving mesh methods. Phase field (PF) method is another approach that has emerged as a powerful tool for simulating problems with moving interface. The PF models represent the moving interface by means of a scalar field variable, the phase field variable, which differentiates between different phases. In galvanic corrosion, for example, two different phases are realized: solid material and electrolyte. Different values are assigned to the phase field variable to distinguish the solid material from the electrolyte. Phase field formulation is based on minimization of an energy functional with respect to admissible driving forces, e.g. temperature in solidification [17], concentration of transport species in corrosion [18], and electrochemical over-potential in electrodeposition [19]. In the past, numerous PF models have been developed for simulation of various phenomena such as bainitic phase transformation in steel [20], solidification [17], and phase transformation in Eutectic alloy [21].

Although PF method has been used for simulation of various physical phenomena with complex moving boundaries, only recently its application for simulating corrosion degradation has been highlighted [18]. Wen *et al*. [18] were the first to propose a PF model for solid-state phase transformation that is induced by diffusion of chemical species from an external source. They demonstrated the capability of their PF model in predicting phase transformation that occurs along the depth of a material during oxidation, sulfidation or corrosion processes. A similar approach was developed by Abubakar *et al*. [22, 23] to approximate microstructural evolution and distribution of transformation induced stress as a result of $V_2O_5$ hot corrosion. In all cited literature above, the Kim-Kim-Suzuki (KKS) PF model was used to represent the total free energy as a weighted sum of free energies of the coexisting solid and liquid phases [24]. There are also other attempts that use PF method for modeling electrochemical processes. For example, Guyer *et al*. [25] used a PF approach to predict the Butler-Volmer electrode reaction kinetics in an electrochemical process. They employed a linear model for phase field variable and showed that the nonlinear Butler-Volmer type of kinetics can be obtained from the space charge double layer close to electrode-electrolyte interface. Okajima *et al*. [26] developed a PF model in which the diffusional mobility coefficient exponentially depends on the over-potential while the rate of electroplating is still linearly dependent on the thermodynamic driving force. Liang *et al*. [27] developed a nonlinear PF model for predicting interface motion in electroplating processes. In their model, the rate of change of phase field variable (thus the interface motion) has a nonlinear relationship with the thermodynamic driving force.

In this paper, we present a PF model for simulating the activation-controlled corrosion with electrode-electrolyte evolution kinetics. We acknowledge that our present model is based on the PF model developed by Liang *et al*. [19, 27] which is modified to represent a galvanic corrosion model. The PF model proposed by Liang *et al*. [19, 27] is employed to describe the current-potential kinetics through a nonlinear correlation between electrode reaction and the rate of change of the phase field variable. The evolution of electrode-electrolyte interface represents the surface recession in galvanic corrosion and is represented by a continuous transition of the phase field variable. The governing equations for electrochemistry along with an evolution equation for the phase field variable form a system of coupled partial differential equations (PDEs). Solution of this set of PDEs yields the electrolyte potential distribution as well as interface evolution. Consequently, the governing equations are solved by means of standard finite element methods without resorting to special treatments for re-meshing or nodal enrichment. To demonstrate the capabilities of the model, we study two benchmark problems. In the first problem, we simulate galvanic corrosion for Magnesium alloy (AE44)–mild steel couple exposed to 5% NaCl medium. In the second problem, we simulate crevice corrosion degradation for Nickel in 1N sulfuric acid for active to passive and passive to active portion of polarization curve. Simulation results are compared with a sharp interface approach as well as available experimental data. A fairly good agreement is observed between the present work and available data. In Appendix A, we show that regardless of the form of current-potential kinetic function the sharp interface model is recovered as a limit of phase field equations in asymptotic analysis.

## 2. Phase field and charge conservation equations

In the present PF model phases (electrode or electrolyte) are represented by different values of phase field variable, $\xi$, which varies between 0 and 1. In this study, $\xi = 1$ and 0 represent electrode (i.e., metal) and electrolyte phases, respectively. The electrode-electrolyte interface is defined by the region $0 < \xi < 1$. Figure 1 shows a schematic of electrode-electrolyte domain with phase field variable ranging between 0 and 1.

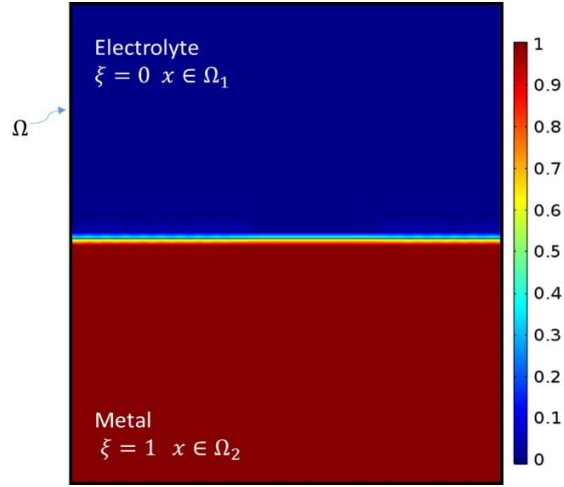

**Figure 1:** Phase field variable in metal, electrolyte and interface regions.

The anodic reaction on metal surface can be written as:

$$M \rightarrow M^{z^+} + ze^- \qquad (1)$$

where $z$ is the number of electrons involved in anodic reaction. Assuming that the electrolyte is well mixed, and no over-potential drop exists due to concentration gradient, the thermodynamic driving force is expressed as:

$$\Delta G = zF(\phi - \phi_{0_a}) = zF\eta \qquad (2)$$

where $G$ is the free energy of the electrode, $F$ is the Faraday constant, $\phi$ is the electrochemical potential, $\phi_{0_a}$ is the corrosion potential of anodic reaction, and $\eta$ is the electrochemical over-potential. Total energy of the electrode-electrolyte system, $\Psi$, consists of corrosion free energy density associated with anodic reaction and gradient energy of electrode-electrolyte interface:

$$\Psi = \frac{1}{v_m}\int_V [e(\xi,\eta) + \frac{1}{2}\epsilon^2(\nabla\xi)^2] dV \qquad (3)$$

The first term in the integral of Eq. (3) represents bulk free energy of the electrode-electrolyte system,

$$e(\xi,\eta) = h(1-\xi)\Delta G + \frac{1}{a}g(1-\xi) \qquad (4a)$$
$$h(\xi) = -2(\xi)^3 + 3(\xi)^2 \qquad (4b)$$
$$g(\xi) = \xi^2(1-\xi)^2 \qquad (4c)$$

where $h(\xi)$ is an interpolation function, $g(\xi)$ is a double-well free energy function and $a$ is the double-well potential coefficient. Note $h(1-\xi)$ or $g(1-\xi)$ in Eq. (4a) means that one needs to replace $\xi$ with $1-\xi$ in Eq. (4b) or (4c). The second term in the integral of Eq. (3) includes gradient of $\xi$ and defines kinetic energy stored in the interface region between electrode $\xi = 1$ and electrolyte $\xi = 0$. The gradient coefficient, $\epsilon$, weights the two terms in Eq. (3) thus controls the interface thickness.

Employing the nonlinear model from Liang *et al.* [27], evolution of the phase field variable can be written as:

$$\alpha\epsilon^2 \frac{\partial \xi}{\partial t} = \frac{\delta \Sigma}{\delta \xi} - L_\eta h'(1-\xi)f(\eta) \qquad (5)$$

where $\alpha$ is the interface mobility, $L_\eta$ is a tunable constant and $\Sigma$ is the total interfacial free energy of the electrode-electrolyte system given by:

$$\Sigma = \frac{1}{v_m}\int_V [\frac{1}{a}g(\xi) + \frac{1}{2}\epsilon^2(\nabla\xi)^2] dV \qquad (6)$$

The $f(\eta)$ in Eq. (5) represents the current-potential kinetic function of the electrode. In particular case of the Butler-Volmer electrode reaction kinetics we have:

$$f(\eta) = i_{0_a}[\exp\left(\frac{\alpha_a zF\eta}{RT}\right) - exp\left(-\frac{\beta_a zF\eta}{RT}\right)] \qquad (7)$$

where $\alpha_a$ and $\beta_a$ are Tafel constants satisfying the relation $\alpha_a + \beta_a = 1$.

The first and second terms on the right hand side of Eq. (5) signify contributions of the interfacial energy and electrode reaction to the evolution of the phase field variable, respectively. The nonlinear dependency of the anodic reaction rate on the over-potential follows the Butler-Volmer kinetics. At low over-potentials, the well-known Allen-Cahn equation [29] can be recovered from Eq. (5) as shown in the work of Liang *et al.* [27].

The following charge conservation equation is solved along with Eq. (5) to obtain the distribution of electric potential:

$$\nabla \cdot \left(\sigma(\xi)\nabla(\eta)\right) = 0 \qquad (8)$$

where the electric conductivity is a function of phase field variable as:

$$\sigma(\xi) = \sigma_0\, h(1-\xi) = -2(1-\xi)^3 + 3(1-\xi)^2 \qquad (9)$$

Expansion of Eq. (8) results in the following expression:

$$-\sigma_0 \frac{\partial h(\xi)}{\partial \xi}\nabla\xi \cdot \nabla(\eta) + \sigma_0 h(1-\xi)\nabla^2(\eta) = 0 \qquad (10)$$

Equation (10) implies that the Laplace equation for charge transfer is integrated with a source term which is the first term on the left hand side of Eq. (10). This source term stems from the generation of anodic current at the interface. We will later show, in Eq. (14), that the source term is proportional to the rate of change of the phase field variable. This proportionality can be obtained using the relation between the interface advection equation, Eq. (11a), and Faraday's law, Eq. (11b), as:

$$\frac{1}{|\nabla \xi|}\left(\frac{\partial \xi}{\partial t} + \vec{v}\cdot\nabla\xi\right) = 0 \qquad (11a)$$

$$\vec{v}\cdot\vec{n} = \frac{M}{\rho F z} i_a \qquad (11b)$$

Using the interface normal vector defined based on the phase field variable, $\vec{n} = \nabla\xi/|\nabla\xi|$, the anodic current at the interface can be obtained in terms of the gradient of the phase field variable:

$$i_a = -\sigma_0 \nabla\eta \cdot \vec{n} = -\sigma_0 \nabla\eta \frac{\nabla\xi}{|\nabla\xi|} \qquad (12)$$

By substituting Eq. (12) into Eq. (11b) and combining the resultant with Eq. (11a), the source term in Eq. (10) can be written in terms of the rate of change of the phase field variable as:

$$\sigma_0 \nabla\eta \nabla\xi = \frac{\partial \xi}{\partial t}\frac{\rho F z}{M} \qquad (13)$$

Substitution of Eq. (13) into Eq. (10) results in Eq. (14) which underlies the relation between the source term and the rate of change of the phase field variable.

$$\sigma(\xi)\nabla^2(\eta) = \frac{\partial \xi}{\partial t}\frac{\rho F z}{M} h'(\xi) \qquad (14)$$

Equation (14) along with Eq. (5) should be solved simultaneously to obtain the distribution of electrolyte over-potential, $\eta$, and the phase field variable, $\xi$. To solve these equations the false transient method is adopted [30]. This method transforms the elliptic PDE in Eq. (14) to a parabolic one by inserting a fictitious transient term on the right hand side of Eq. (14) with modified constant coefficients $\gamma_\xi$ and $\gamma_\eta$. Therefore, equations for the phase field evolution and the electrolyte over-potential can be written as:

$$\gamma_\xi \alpha \epsilon^2 \frac{\partial \xi}{\partial t} = \frac{1}{a} g'(\xi) + \epsilon^2 \nabla^2 \eta + L_\eta h'(\xi) f(\eta) \qquad (15)$$

$$\frac{\gamma_\eta d\eta}{dt} = \sigma(\xi)\nabla^2(\eta) - lh'(\xi)\frac{\partial \xi}{\partial t} \qquad (16)$$

Invoking the false transient method requires using initial conditions at time $t_0$ both for the phase field variable, $\xi$, and the electrochemical over-potential, $\eta$, as:

$\xi_{electrolyte}(t_0) = 0$     (17a)

$\xi_{electrode}(t_0) = 1$     (17b)

$\eta(t_0) = 0$     (17c)

where $l = \frac{\rho F z}{M}$ can be interpreted as a latent electricity coefficient. Results of using Eqs. (15) and (16) for two benchmark problems are presented in Section 3. In Appendix A we show that the sharp interface model is recovered as the limiting case of the proposed PF model.

## 3. Results

*Case study 1: Galvanic corrosion of AE44(Mg)-Mild steel couple*

Corrosion of a galvanic couple, Magnesium alloy (AE44)–mild steel, exposed to 5% NaCl medium is simulated using PF model. Results of the present work is compared with available experimental data as well as modeling works of Sun *et al*. [31] and Deshpande [6,32,33] which are based on Arbitrary Lagrangian Eulerian (ALE) method. The ALE method is a moving mesh technique in which spatial frame has coordinates that are moving with time. Readers interested in details of the ALE method may refer to [6]. We implemented the PF and ALE methods into a finite element code using COMSOL. The ALE method, as a sharp interface approach, is used as a reference solution. We further compare the PF model predictions with available experimental data.

Model results are obtained by solving Eqs. (15) and (16) over the electrode-electrolyte domain enclosed by solid lines shown in Figure 2. Polarization data for AE44 (Mg) and mild steel are adopted from [6] and are shown in Figure 3. The Butler-Volmer relation is used to represent the electrode kinetics. Boundary condition for the cathode surface, numbered as 4 and 5 in Figure 2, is expressed as:

$$\nabla_n \eta = i_c = (1 - \xi)i_{0_c}[\exp\left(\frac{\alpha_c zF(\eta + \phi_{0_a} - \phi_{0_c})}{RT}\right) - \exp\left(-\frac{\beta_c zF(\eta + \phi_{0_a} - \phi_{0_c})}{RT}\right)] \quad (18)$$

Note that the cathode region, shown in Figure 2, is not part of the computational domain while boundary 4 and 5 are included. Zero flux boundary conditions are applied to all other boundaries:

$\nabla_n \eta = 0$ (19a)
$\nabla_n \xi = 0$ (19b)

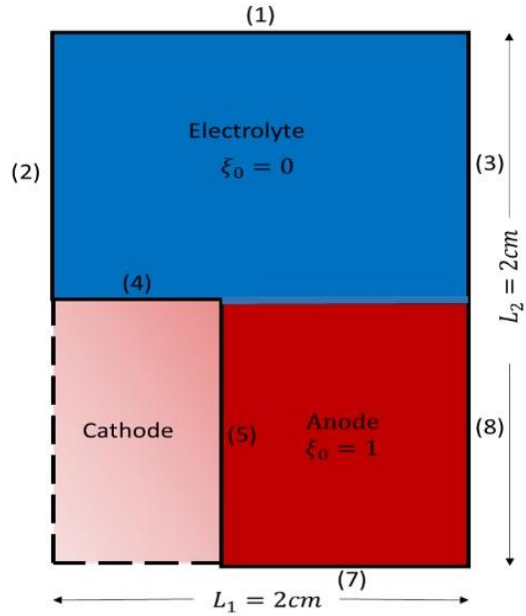

**Figure 2:** Schematic of the computational domain used for PF modeling of galvanic corrosion in AE44 (Mg) - mild steel coupling. Please note that colors are used in this figure to distinguish different regions and not representing any quantity.

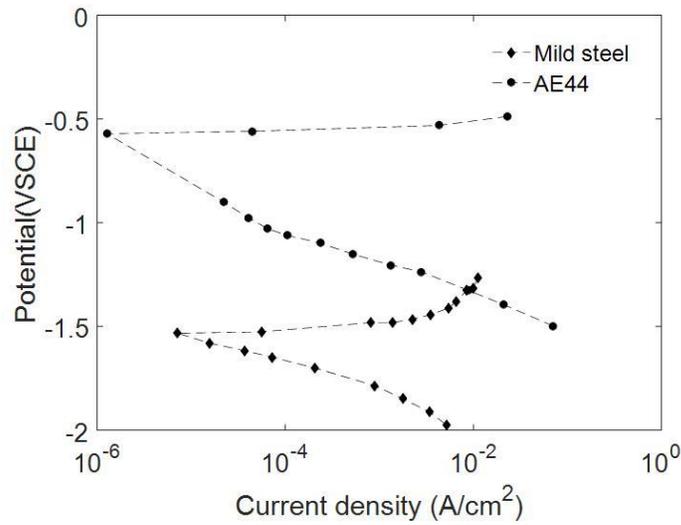

**Figure 3:** Polarization curves for AE44-mild steel galvanic couple (obtained from [6]).

The PF model predictions for evolution of surface morphology and potential distribution over 12 days of continuous exposure to electrolyte are shown in Figure 4. It is to be noted that X and Y in Figure 4 are coordinates of the computational domain in *x* and *y* direction, respectively. For example, the cathode-anode interface is located at X = 10 mm. The colorbar in this figure shows the variation of electric potential with unit of V(SCE). It can be seen that the electric potential varies from -1.28 V(SCE) over the cathode to -1.39 V(SCE) over the anode. As expected, the anodic dissolution is higher close to the cathode-anode interface and decreases along the distance away from the interface. The maximum surface recession has been predicted to be about 2.5 mm for 5 days of exposure. This value for 8 days and 12 days of exposure has been 4.2 mm and 6 mm, respectively.

To make a quantitative comparison, corrosion current density after 3 days of immersion predicted by the PF model is compared to that obtained from the sharp interface model solving Eq. (11b) and Eq. (A2) from Appendix A. Figure 5(a) shows the solution domain for the sharp interface model. The sharp interface model uses the ALE technique to obtain the evolution of surface recession. Figure 5(b) shows the distribution of current density on anode and cathode surfaces obtained from the present PF model and the ALE approach presented in [6]. Results show a good agreement between the two methods. Figure 5(c) shows the corrosion depth predicted by PF model, ALE approach and those obtained from immersion experiments reported in [6]. Figure 5(d) shows qualitative comparison of corrosion depth between PF model prediction and experimental data. The results of model prediction agree fairly well with experimental data except at regions close to cathode-anode interface and right edge of the domain, as seen in Figure 5(c), i.e., X < 11 mm and X > 19 mm. This discrepancy may be due to the boundary conditions imposed by Eqs. (18) and (19) and the assumption that the electrolyte solution is well mixed with no concentration gradient exists in the electrolyte solution. In the current analysis, we did not account for the effect of variations in electrolyte composition and ionic strength on the electrochemical process. We acknowledge that the species concentrations may vary spatially due to finite size of the computational domain and diffusion of species may have significant effect on corrosion rate at local sites. Interestingly, an increase in the depth of corrosion attack near the right edge of galvanic couple, i.e., X > 19 mm, is observed in experimental results as seen in Figs. 5(c) and 5(d). This behavior is not captured with either ALE or PF method. This behavior which is referred to as "edge effects" occurs because of enhanced mass transport at electrode-insulator interface [34]. According to Trinh *et al*. [34] the corrosion process at the electrode−insulator interface can significantly be influenced by a chemically inert material surrounding a galvanic couple. Although Deshpande [6, 33] does not explicitly discuss the presence of an insulator surrounding the galvanic couple in his tests, Trinh *et al*. [34] explain the edge effect in relation to Deshpande's work. This effect is beyond the scope of present study and will be considered in future work.

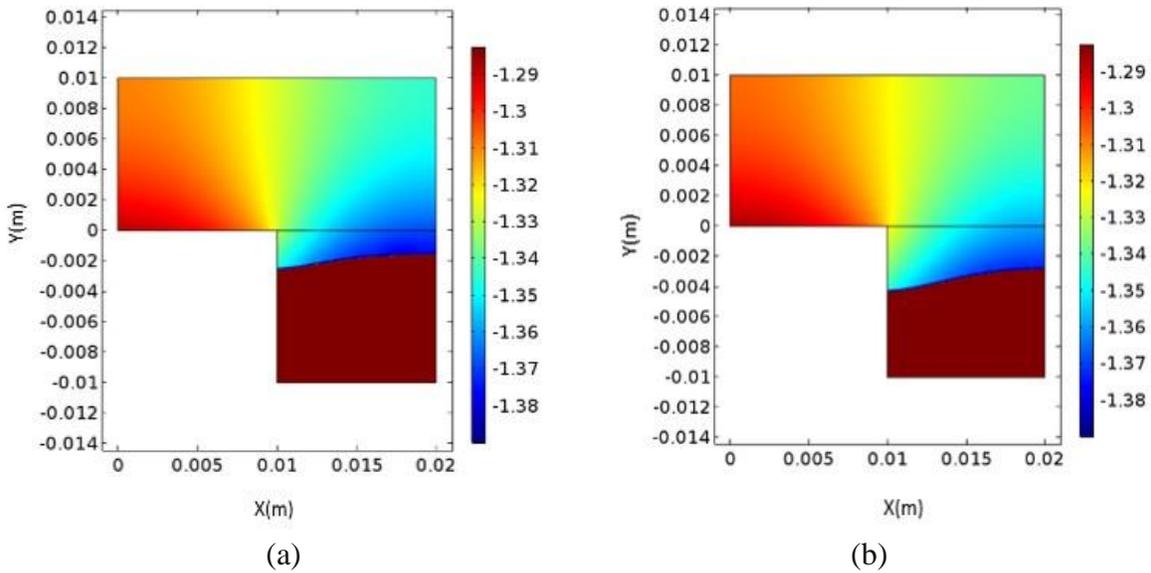

(a)     (b)

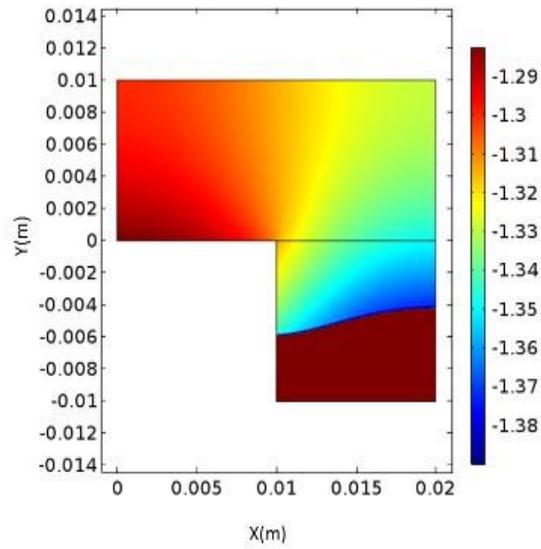

(c)

**Figure 4:** Predicted surface morphology and corresponding electric potential distribution for AE44–mild steel galvanic couple over a) 5 days, b) 8 days and c) 12 days of continuous exposure to 5% NaCl. For interpretation of color in this figure, the reader is referred to the web version of this article.

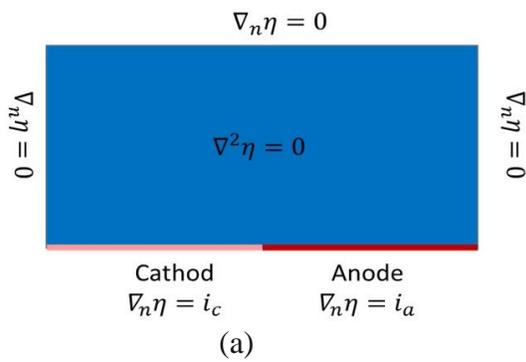

(a)

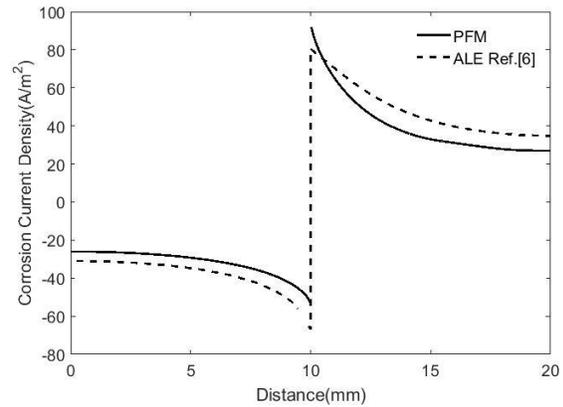

(b)

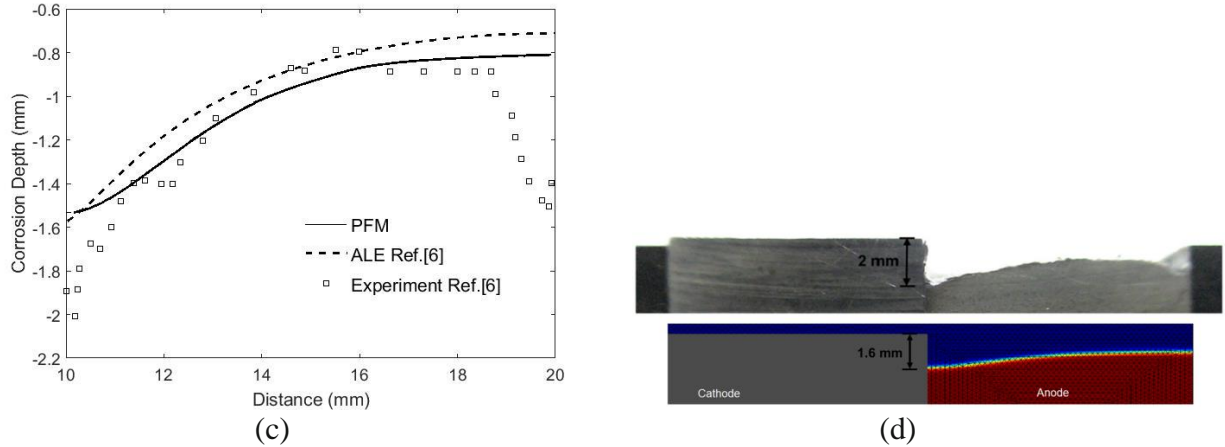

(c)  (d)

**Figure 5:** AE44–mild steel galvanic couple corrosion: a) schematic of boundary conditions used for ALE method (sharp interface), b) current density after 3 days of exposure to 5% NaCl, using PF model and ALE technique, c) corrosion depth profile predicted by PF and ALE models, and those obtained from experimental data [6], d) qualitative comparison of corrosion depth between PF model prediction and experimental data of [6].

*Case study 2: Crevice corrosion*

We implemented the PF model for 2D modeling of crevice corrosion. Figure 6 shows a schematic of the crevice geometry including metal and electrolyte domains. We simplify the problem by assuming that electrolyte in bulk and crevice region is well mixed. The potential distribution and electrode deformation are obtained by solving Eqs. A4 and A5 and adapting polarization curve for nickel in sulfuric acid from [35] (see Figure 7). Surface profile and distribution of current density along the crevice are obtained. Results are compared with the work of Brackman *et al*. [36] who used finite volume (FV) formulation with a level set approach to simulate the potential distribution and the surface profile. We further compare the PF model predictions with experimental work of Abdulsalam and Pickering [35], and Lee *et al*. [28]. Figures 8(a) and 8(b) present potential distribution and anodic current density along the crevice after 50 hours of exposure to electrolyte, respectively. Results of the work of Abdulsalam and Pickering [35] and Brackman *et al*. [36] are also plotted for comparison. PF model predictions fairly agree with modeling work of Brackman *et al*. [36] and experimental work of Abdulsalam and Pickering [35]. Profile of the corroded surface is compared with modeling results of Brackman *et al*. [36] and experimental results of Abdulsalam and Pickering [35] in Figure 9. While comparison of PF model prediction for corroded surface is in good agreement with modeling results of Brackman *et al*. [36] and ALE model, there is a poor agreement between experiment and model predictions, particularly at crevice depth of 2-3 mm. This discrepancy is due to some incorrect measurement (e.g., solution homogeneity, polarization behavior and IR drop) made by Abdulsalam and Pickering [35] which is also discussed in [36]. Further, while the current density is sufficiently high at depth of 4 mm, no damage is indicated by experimental results at this depth, which further implies that there might be inaccurate measurements of corrosion depth in the experimental work of Abdulsalam and Pickering [35].

In this case study, current-potential kinetic ranges from passive to active and active to passive, and does not necessarily follows the Butler-Volmer electrode reaction kinetics. It is, however, shown that regardless of the form of current-potential kinetic, $f(\eta)$, PF model is capable of predicting crevice corrosion.

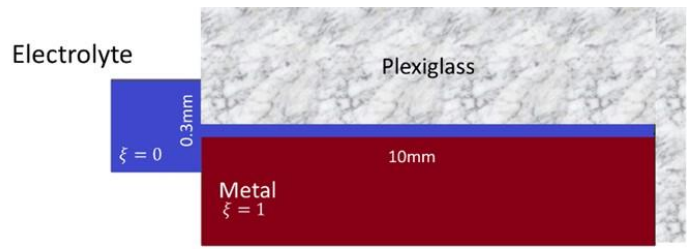

**Figure 6:** Electrode and electrolyte computational domain

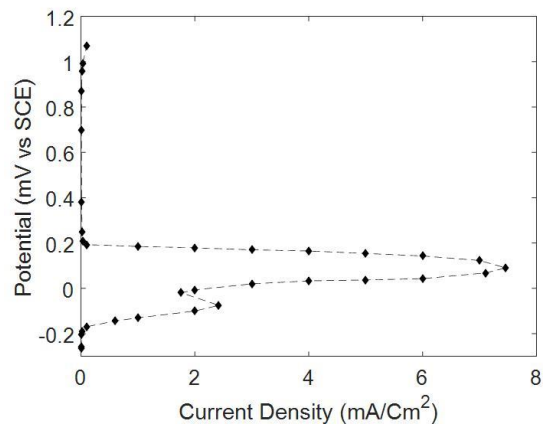

**Figure 7:** Polarization curve or pure nickel in sulfuric acid from Abdulsalam and Pickering [35]

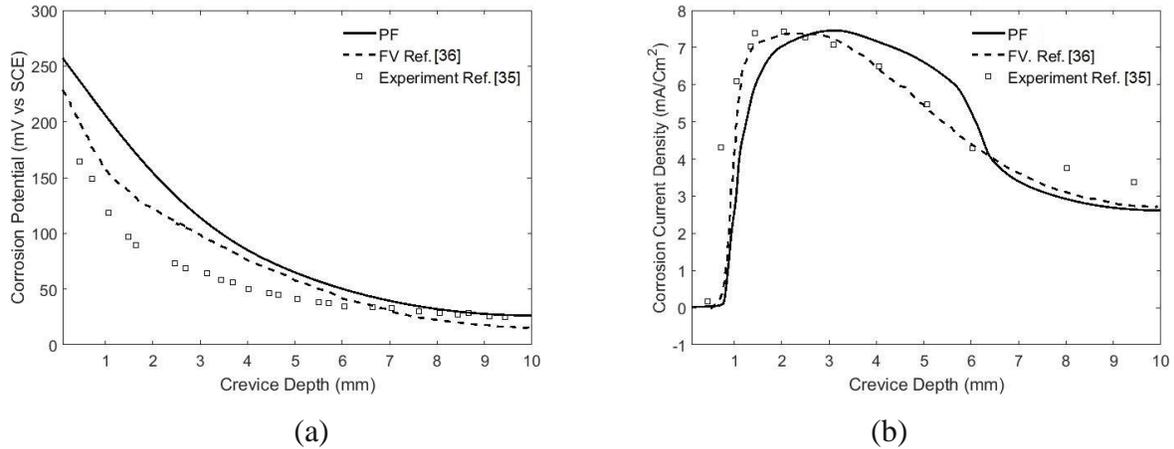

(a)            (b)

**Figure 8:** Distribution of a) corrosion potential and b) corrosion current density obtained from PF model, Abdulsalam and Pickering [35] and Brackmen *et al*. [36]

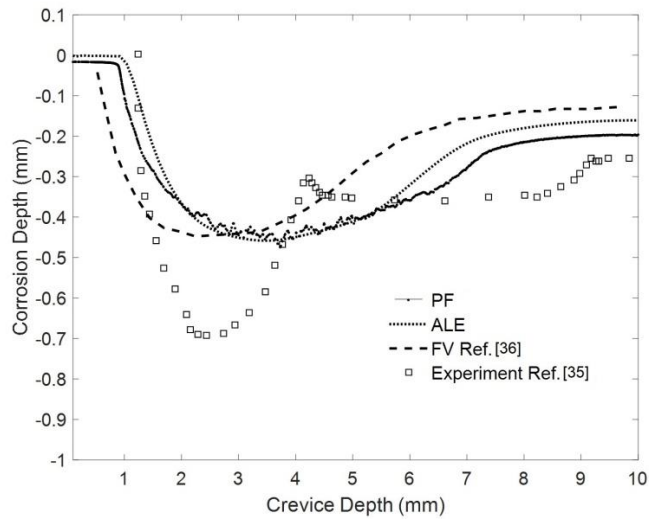

**Figure 9:** Corrosion depth profile obtained from PF model, ALE, Abdulsalam and Pickering [35] and Brackmen *et al*. [36]

### 4. Conclusions

We presented a phase field model for prediction of galvanic corrosion. We acknowledge that our present model is based on the phase field model developed by Liang *et al*. [19, 27] which is modified to represent a galvanic corrosion model. By using the proposed model, evolution of corroded surface was naturally obtained from solving the phase field equations without the need to track the moving interface. We further showed that the classical sharp interface equations of potential and current kinetics can be obtained as a particular limit of the phase field model. Two case studies were presented to demonstrate capability of the model in predicting corrosion damage: galvanic corrosion and crevice corrosion. Results of the phase field model were compared with experimental data as well as sharp interface modeling results. Our future work effort includes implementation of the proposed phase field model for complex corrosion problems such as pitting corrosion which involves complex evolution of the interface multiple pits interact.

**Acknowledgement**

Authors would like to thank Mr. William C. Nickerson from Office of Naval Research, ONR, for financial support provided under grant/contract number SBIR Phase II Contract N68335-16-C-0135.

**Appendix A**

*Asymptotic solution*

We demonstrate that the sharp interface model can be recovered as a limiting case of the proposed PF model. Galvanic corrosion can be considered as a two phase problem with electrolyte and solid material as two distinct phases. The electrolyte and solid phases are shown as region $\Omega_1$ and $\Omega_2$ in Figure 1, respectively. The electrolyte region is defined by:

$$\Omega_1 = \{x \in \Omega, \sigma > 0\} \qquad (A.1)$$

with $\sigma$ being constant over $\Omega_1$ in the sharp interface approach. The over-potential must satisfy the charge transfer equation in $\Omega_1$ as:

$$-\sigma \nabla^2 \eta = 0, \quad x \in \Omega_1 \qquad (A2)$$

Faraday's law presented with Eq. 11b must be also satisfied for $\forall x \in \Gamma$. We follow the approach presented in [17] and show that sharp interface equations represented by Eqs. 11b and A2 are recovered from the phase field equations with $\epsilon, a \to 0$, fixed $\gamma_\xi \alpha$ and $\epsilon a^{-\frac{1}{2}}$. Verification of this proposition can be made by setting:

$$c_1 = \epsilon a^{-1/2}, \epsilon_1 = \epsilon^2, p(\xi) = L_\eta h'(\xi), \alpha' = \gamma_\xi \alpha, q(\xi) = c_1^2 g'(\xi) \qquad (A3)$$

and rewriting Eqs. 15 and 16 as:

$$\alpha' \epsilon_1^2 \frac{\partial \xi}{\partial t} = q(\xi) + \epsilon_1^2 \nabla^2 \eta + \epsilon_1 p(\xi) f(\eta) \qquad (A4)$$

$$\frac{\gamma_{eta} d\eta}{dt} = \sigma(\xi) \nabla^2 (\eta) - lh'(\xi) \frac{\partial \xi}{\partial t} \qquad (A5)$$

We further carry out asymptotic expansions of Eqs. A4 and A5 following the method of matched asymptotic expansions presented in [17]. Using this method the inner and outer solutions can be obtained. We finally conclude that leading interface profiles satisfy the Faraday's law.

Let us define $(r, s)$ as the local coordinate system in a small neighborhood around $\Gamma(t)$, where $r(x, y, t)$ is positive in the direction of $\xi = 1$ and negative in the direction of $\xi = 0$, and $s(x, y, t)$ measures the arc length from a fixed point. In the neighborhood of $\Gamma(t)$ we have:

$$|\nabla r| = 1; \qquad \nabla^2 r = k \qquad (A6)$$

We now seek solutions of Eqs. A4 and A5 in the inner region where $\xi$ varies rapidly and in the outer region where $\xi$ varies slowly in bulk phases far from interface.

Expansion of variables in the outer region results in:

$$\eta(x, y, t, \epsilon_1) = \eta^0(x, y, t) + \epsilon_1 \eta^1(x, y, t) + \epsilon_1^2 \eta^2(x, y, t) + \cdots \qquad (A7)$$

$$\xi(x, y, t, \epsilon_1) = \xi^0(x, y, t) + \epsilon_1 \xi^1(x, y, t) + \epsilon_1^2 \xi^2(x, y, t) + \cdots \qquad (A8)$$

$$r(x, y, t, \epsilon_1) = r^0(x, y, t) + \epsilon_1 r^1(x, y, t) + \epsilon_1^2 r^2(x, y, t) + \cdots \qquad (A9)$$

$$s(x, y, t, \epsilon_1) = s^0(x, y, t) + \epsilon_1 s^1(x, y, t) + \epsilon_1^2 s^2(x, y, t) + \cdots \qquad (A10)$$

By defining a new variable, $w$, as:

$$w = \frac{r}{\epsilon_1} \qquad (A11)$$

the inner expansions can be obtained as:

$$\eta(x,y,t,\epsilon_1) = E^0(w,s,t,\epsilon_1) + \epsilon_1 E^1(w,s,t,\epsilon_1) + \epsilon_1^2 E^2(w,s,t,\epsilon_1) + \cdots \quad \text{(A12)}$$

$$\xi(x,y,t,\epsilon_1) = X^0(w,s,t,\epsilon_1) + \epsilon_1 X^1(w,s,t,\epsilon_1) + \epsilon_1^2 X^2(w,s,t,\epsilon_1) + \cdots \quad \text{(A13)}$$

*The outer expansion:*

A new set of equations for Eqs. A4 and A5 is obtained by using Eqs. A7- A10 and matching formal order.

For $O(1)$ we get:

$$q(\xi^0) = 0 \quad \text{(A14)}$$

$$\gamma_{eta}\eta_t^0 = \sigma(\xi^0)\nabla^2(\eta^0) - lh'(\xi^0)\xi^0 \quad \text{(A15)}$$

Equation A14 has solutions of 0 and 1 which indicates that for $r \neq 0$ or the bulk of electrolyte where $h(1-\xi) = 1$ and $h'(\xi) = 0$, Eq. A15 reduces to charge transfer equation, Eq. A2, when no transient coefficient is considered. Since balances of higher order $\epsilon_1$ are not informative their formulation is eliminated in this section.

*The inner expansion:*

Using $(r,s)$ coordinate system, the Laplacian and the time derivative can be written as:

$$\nabla \xi = \xi_{rr} + \nabla r \xi_r + |\nabla s|^2 \xi_{ss} + \nabla^2 s \xi_s \quad \text{(A16)}$$

$$\xi_t(x,y) = \xi_t(r,s) + r_t \xi_r + s_t \xi_s \quad \text{(A17)}$$

Invoking Eqs. A12 and A13, a new set of equations for Eqs. A4 and A5 is obtained as:

$$X_{ww} + q(X) + \epsilon_1(\nabla^2 r\, X_w - \alpha' r_t X_w + p(X)f(E)) +$$
$$\epsilon_1^2(X_{ss}|\nabla s|^2 + X_s \nabla^2 s - \alpha' X_t - \alpha' s_t X_s) \quad \text{(A18)}$$

$$X_{ww} + q(X) + \epsilon_1(\nabla^2 r\, X_w - \alpha' r_t X_w + p(X)f(E)) +$$
$$\epsilon_1^2(X_{ss}|\nabla s|^2 + X_s \nabla^2 s - \alpha' X_t - \alpha' s_t X_s) \quad \text{(A19)}$$

For $O(1)$ we obtain the followings:

$$X^0_{ww} + q(X^0) = 0 \qquad (A20)$$

$$\sigma(X^0)E^0_{ww} = 0 \qquad (A21)$$

Since in the inner region $\sigma(X^0) \neq 0$, Eq. A21 reduces to $E^0 = aw + b$. Using the matching condition method, we obtain:

$$E^0(\pm\infty, t) = E^0(\Gamma^0_\pm, t) \qquad (A22)$$

Equation A22 is only valid when $a = 0$, otherwise $E^0$ would be unbounded at the interface. This implies:

$$E^0 = b(s, t) \qquad (A23)$$

By using the same matching condition, we have:

$$X^0(+\infty, t) = X^0(\Gamma^0_+, t) = 1 \qquad (A24)$$

$$X^0(-\infty, t) = X^0(\Gamma^0_-, t) = 0 \qquad (A25)$$

The $O(\epsilon_1)$ balance in Eq. A21 yields:

$$\sigma(X^0)(E^1_{ww}) = -lh'(X^0)r^0_t X^0_w \qquad (A26)$$

Integrating Eq. A26 results in:

$$\sigma(X^0)(E^1_w) = lh(X^0)r^0_t X^0 + c_1(s, t) \qquad (A27)$$

Note that $E^0_w = 0$. Using the matching condition and differentiating with respect to $w$, we obtain:

$$\lim_{w \to \pm\infty}(E^1(w,t)) = -\eta^0_r(\Gamma^0_\pm, t) \qquad (A28)$$

Equation A28 along with Eq. A27 and boundary conditions of Eqs. A24 and A25 result in the following interface conditions:

$$\sigma(X^0)(\eta^0_r|_{\Gamma_+}) = lh(X^0)|_{\Gamma_+} r^0_t + c_1(s, t) \qquad (A29)$$

$$\sigma(X^0)(\eta^0_r|_{\Gamma_-}) = c_1(s, t) \qquad (A30)$$

Note that the value of normal velocity, $v$, is given by $r_t$, $h(X^0)|_{\Gamma_+} = 1$. By subtracting Eq. A29 from Eq. A30, the electrode-electrolyte interface velocity can be recovered as:

$$\sigma(X^0)\left(\eta_r^0|_{\Gamma_\pm}\right) = lv \qquad (A31)$$

which is the same as the sharp interface model. Therefore, we have shown that the interface boundary condition can be extracted from the phase field equations (Eq. 11b). It was also demonstrated that the result is independent of the form of current-potential kinetic function, $f(\eta)$, in particular, Butler-Volmer electrode reaction kinetic function.